\documentclass[twocolumn,showpacs,preprintnumbers,amsmath,amssymb]{revtex4-1}
\usepackage{amsfonts}
\usepackage{amsmath}
\usepackage{graphicx}
\usepackage{dcolumn}
\usepackage{bm}
\usepackage{overpic}
\usepackage{booktabs}
\usepackage{color}

\begin{document}
\title{Ranking spreaders by decomposing complex networks}
\author{An Zeng}\email{an.zeng@unifr.ch}
\author{Cheng-Jun Zhang}
\affiliation{Department of Physics, University of Fribourg, Chemin du Mus\'{e}e 3, CH-1700 Fribourg, Switzerland}
\date{\today}

\begin{abstract}
Ranking the nodes' ability for spreading in networks is a fundamental problem which relates to many real applications such as information and disease control. In the previous literatures, a network decomposition procedure called k-shell method has been shown to effectively identify the most influential spreaders. In this paper, we find that the k-shell method have some limitations when it is used to rank all the nodes in the network. We also find that these limitations are due to considering only the links between the remaining nodes (residual degree) while entirely ignoring all the links connecting to the removed nodes (exhausted degree) when decomposing the networks. Accordingly, we propose a mixed degree decomposition (MDD) procedure in which both the residual degree and the exhausted degree are considered. By simulating the epidemic process on the real networks, we show that the MDD method can outperform the k-shell and the degree methods in ranking spreaders. Finally, the influence of the network structure on the performance of the MDD method is discussed.
\end{abstract}
\maketitle

\section{Introduction}

Spreading is an important process widely existing in various fields including physics, chemistry, medical science, biology and sociology~\cite{RMP801275}. For example, the reaction diffusion processes~\cite{NP3276}, pandemics~\cite{PRL863200}, cascading failures in electric power grids~\cite{PRL93098701,Nature4641025} and information dissemination~\cite{EPL8838005} can be naturally characterized by the framework of spreading. In particular, spreading in complex networks has been intensively studied in the past decade. Many studies have revealed that the spreading process is strongly influenced by the network topologies~\cite{PRL89108701,PRL90028701,PRL97088701,PRL105218701}. With the understanding of spreading pathways on networks, many methods have been developed to manipulate network structure to control the spreading threshold~\cite{PRE85015101,EPL9518005}. Moreover, in order to avoid the wide propagation of the disease, various efficient immunization strategies were also proposed~\cite{PRL91247901,PRL101058701}.

Though lots of former works are dedicated to understand and control the spreading process in a macroscopic sense, recently more and more attentions have been paid to microscopically study the \emph{spreadability} for each node, i.e., how many nodes will finally be covered when the spreading originates from this single node~\cite{NP6888,PhysicaA3911777,arxiv12020024}. The knowledge of node spreadability is crucial for developing efficient methods to either decelerate spreading in the case of diseases, or speed up spreading in the case of information flow. Moreover, it can be helpful for identifying the initial spreader of certain disease or information~\cite{PRE84056105}. Though the most connected nodes (hubs) and the nodes with high betweenness centrality are commonly believed to be the most influential spreaders in networks, the k-shell (also called k-core) method is found to perform better in identifying the best individual spreaders~\cite{NP6888,PNAS1074491}. The k-shell method starts by removing all nodes with one connection only (with their links), until no more such nodes remain, and assign them to the $1$-shell. For each remaining node, the number of links connecting to the other remaining nodes is called its \emph{residual degree} and the number of links connecting to the removed nodes is called its \emph{exhausted degree}. After assigning the $1$-shell, all nodes with residual degree $2$ are recursively removed and the $2$-shell are created. This procedure continues as the residual degree increases until all nodes in the nodes have been assigned to one of the shells. The nodes with high k-shell value tend to locate in the center of the network and the spreading starting from each of these nodes are likely to widely cover the network. Actually, similar idea has also been applied to assign direction to the link of undirected networks and significant improvement in synchronizability can be achieved~\cite{PRL103228702,PRE83045101}.

We find, however, that the k-shell method has several limitations when it is used to rank the spreadability for all the nodes. First, it assigns many nodes with the same rank even though they perform entirely different in spreading. The extreme examples are the tree network~\cite{EPL8748002} and Barabasi-Albert network~\cite{Science286509} in which the k-shell method assigns every node with the same shell. Second, the assigned shell by this method cannot correctly reflect the real spreadability of nodes in some cases. For instance, if a hub $i$ connects a large number of tree-like branches, the k-shell method will still assign the hub with $k_{s}=1$. However, if a node $j$ with low degree forms only one triangle with other nodes, it will have $k_{s}=2$. Apparently, node $i$ should performs far better than $j$ as a spreader in reality. These two limitations make $k_{s}$ unable to be used to accurately rank the spreadability of nodes.

Actually, the above-mentioned limitations for k-shell method is due to entirely ignoring all the links of the removed nodes when decomposing the networks. In this paper, we propose the so-called mixed degree decomposition (MDD) procedure in which both the residual degree and the exhausted degree are taken into account. By simulating the epidemic process on the real networks, we show that the MDD performs more accurately than the k-shell and the degree methods in ranking the spreadability for nodes. Finally, we discuss how the structure of real network affects the performs of the MDD method.


\section{Method}

The k-shell method is a dynamical network decomposition procedure in which the residual degree of nodes should be updated in each step. During the decomposition all the information of the removed nodes are dropped so that this method assumes that the remaining nodes are homogeneously connecting to the removed nodes. In other words, if the virus or information reaches a certain layer, each node in this layer is assumed to spread the virus/information to the same number of nodes in the lower layers, which is not true in reality. If a node in a low layer connects to a big branch of removed nodes, not only this node should be ranked higher than the other nodes in this layer, but also it may have stronger spreadability than some nodes in the higher layers.

The analysis above requires us to take the information of removed nodes into consideration during the decomposition procedure. For a node $i$, we denote the residual degree (number of links connecting to the remaining nodes) and the exhausted degree (number of links connecting to the removed nodes) as $k^{r}_{i}$ and $k^{e}_{i}$ respectively. To achieve a more accurate ranking for node spreadability, we propose a Mixed Degree Decomposition (MDD) procedure in which the nodes are removed in each step according to the mixed degree
\begin{equation}
k^{m}=k^{r}+\lambda*k^{e}_{i},
\end{equation}
where $\lambda$ is a tunable parameter between $0$ and $1$. The detailed decomposition is done with the following procedure:

\begin{enumerate}
\item Initially, $k^{m}$ of each node is equal to $k^{r}$ since there is no removed node in the network.
\item Remove all the nodes with the smallest $k^{m}$ (denoted as $M$) and assign them to the $M$-shell.
\item Update $k^{m}$ of all the remaining nodes by $k^{m}=k^{r}+\lambda*k^{e}_{i}$. Then, remove all the nodes with $k^{m}$ smaller than or equal to $M$ and assign them to the $M$-shell too. This step is recursively carried on until $k^{m}$ of all remaining nodes are larger than $M$.
\item Repeat step $2$ and $3$ as $M$ value increases until all nodes in the network have been assigned to one of the shells.
\end{enumerate}

Apparently, when $\lambda=0$, the MDD method returns to the k-shell method in ref.~\cite{NP6888,PNAS1074491}. When $\lambda=1$, the MDD method is equivalent to the degree centrality. Different from the original k-shell method, note that the shell values in MDD method are no longer integer since $k^{m}$ can be decimal when $\lambda$ is between $0$ and $1$. To better illustrate the procedure of MDD, a simple example is shown in Fig. 1 in which parameter $\lambda$ for the MDD method is set as $0.7$.

\begin{figure}
  \center
  \includegraphics[width=9cm]{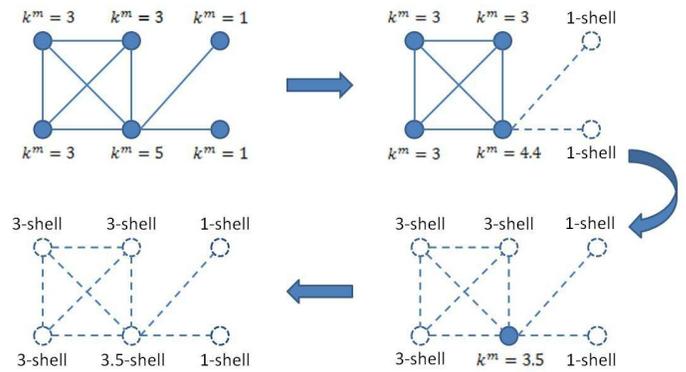}
\caption{(Color online) A simple example to illustrate the procedure of the Mixed Degree Decomposition (MDD). The nodes and links with dashed line represent respectively the removed nodes and exhausted links. Here, the parameter $\lambda$ in MDD is set as $0.7$.}
\label{fig1}
\end{figure}

\section{Result}

To validate the effectiveness of the MDD method, we then apply it to real networks which include social and nonsocial networks. Social networks are: Dolphins (friendship)~\cite{dolphins}, Jazz (musical collaboration)~\cite{jazz}, Netsci (collaboration network of network scientists)~\cite{netcoauthor_word}, Email (communication)~\cite{email}, HEP (collaboration network of high-energy physicists)~\cite{HEP_astrophys_condmatt}, PGP (an encrypted communication network)~\cite{PGP}, Astro phys (collaboration network of astrophysics scientists )~\cite{HEP_astrophys_condmatt}, Cond matt (collaboration network of condensed matter scientists)~\cite{HEP_astrophys_condmatt}. Nonsocial networks are: Word (adjacency relation in English text)~\cite{netcoauthor_word}, E. coli (metabolic)~\cite{Ecoli}, C. elegans (neural)~\cite{Celegans}, TAP (yeast protein-protein binding network generated by tandem affinity purification experiments)~\cite{TAP}, Y2H (yeast protein-protein binding network generated using yeast two hybridization)~\cite{Y2H}, Power (connections between power stations)~\cite{Power}, Internet (router level)~\cite{Internet}. To better illustrate the performance of the MDD method, we select four relatively large networks (email, PGP, Astro phys and Cond matt) as examples and show their results by figures throughout the paper. The results of other networks are detailedly reported in Table I.

\begin{figure}
  \center
  \includegraphics[width=9cm]{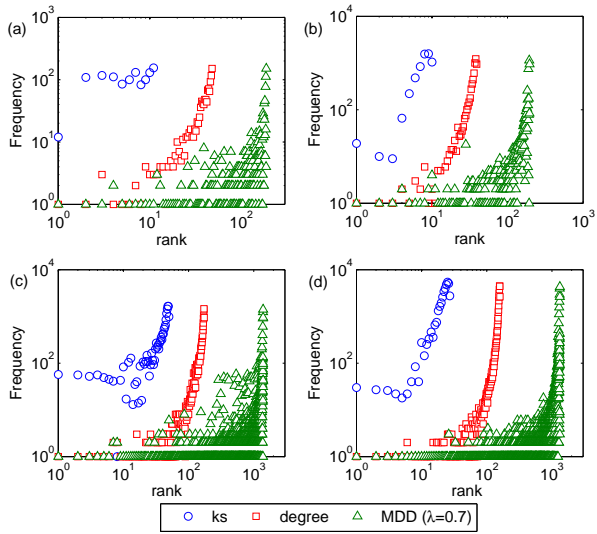}
\caption{(Color online) The frequency of different ranks in k-shell method, degree centrality method and MDD method ($\lambda=0.7$). The networks are: (a) email, (b) PGP, (c) astro phys and (d) condensed matter. }
\label{fig2}
\end{figure}

As we mentioned above, one of the limitation of the original k-shell method is that it is a coarse grained method which assigns many nodes with the same shell (which is equivalent to assigning them with the same rank in spreadability). In Fig. 2, we show the frequency of different ranks in k-shell method, degree centrality method and MDD method. Obviously, the k-shell only has limited number of ranks and the frequency of each rank is quite high, which implies that node differences are not well distinguished in the k-shell method. By using the degree centrality to rank the nodes, larger number of ranks will be obtained. In the MDD method, nodes are more detailedly ranked than the previous two methods and the number of ranks can be even ten times larger than the degree method. More importantly, frequency of the top-rank is almost $1$  which suggests that these nodes are well separated. We also check the performance of MDD method on the tree network and BA model in which k-shell method is not valid, the results show that the MDD can easily detect the difference between nodes.

All the ranking generated by k-shell, degree and MDD methods are obtained by analyzing network topology. In principle, an effective topology-based ranking should be as close as possible to ranking by the real spreading coverage. In this paper, we employ the SIR model~\cite{RMP801275} to simulate the spreading process on networks. The number of final infections resulting from a given initially-infected node $i$ is denoted as its spreadability $s_{i}^{p}$ where $p$ is the infection rate in the SIR model. For all the methods mentioned above, a final ranking will be generated. We therefore use the Kendall's tau rank correlation coefficient ($\tau$) to estimate how the a certain topology-based ranking is correlated to the ranking by the true spreadability $s$ of the nodes. In the most ideal case where $\tau=1$, for each two nodes $i$ and $j$, if $i$ is ranked before $j$ by the topology-based method, $i$ will have stronger spreadability than $j$. In Fig. 3, we show the value of $\tau$ of the k-shell, degree centrality and MDD methods under different $p$. In this paper, we use relatively small values for $p$, namely $p\in (0,0.5]$, so that the infected percentage of the nodes is not so large. In the case of large $p$ values, where spreading can cover almost all the network, the role of individual nodes is no longer important since the final coverage of virus is independent of where it originated from. Interestingly, though the k-shell method is claimed to be able to identify the most influential node, its $\tau$ value is not significantly higher than that of the degree centrality method. Due to the two limitations pointed out in the introduction section, the k-shell method cannot effectively reflect the spreadability of those nodes with low rankings. We again set $\lambda=0.7$ in the MDD as an example and show its $\tau$ value in Fig. 3. As we can see, the MDD outperforms both the k-shell method and the degree centrality method under all the $p$ value we considered.

In order to systematically study how the parameter $\lambda$ affects the performance of the MDD method, we calculate a $\langle\tau\rangle$ by summing all the $\tau$ under different infection rate $p$, namely $\langle\tau\rangle=\sum_{p=0}^{0.5}\tau(p)$. In this way, we can investigate under which $\lambda$ the MDD can achieve the largest $\langle\tau\rangle$, which means the MDD ranking can most accurately reflect the general spreadability of nodes. The related results are shown in Fig. 4. Clearly, neither the k-shell method nor the degree centrality method performs good enough. However, by increasing (decreasing) a little bit of $\lambda$ when $\lambda=0$ ($\lambda=1$), the $\langle\tau\rangle$ can be significantly improved. Moreover, for each network, there is an optimal $\lambda^{*}$ under which the MDD method can achieve an largest $\langle\tau\rangle$. The results for other real networks are reported in Table I. The results show that the optimal $\lambda^*$ universally exists in both social and nonsocial networks.

\begin{figure}
  \center
  \includegraphics[width=9cm]{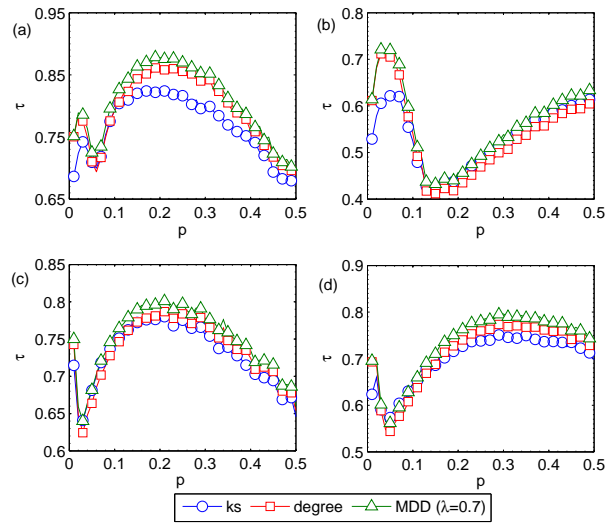}
\caption{(Color online) The value of $\tau$ of the k-shell, degree centrality and MDD methods under different infection rate $p$ in the SIR model. The networks are: (a) email, (b) PGP, (c) astro phys and (d) condensed matter. The results are averaged over $100$ independent realizations. }
\label{fig3}
\end{figure}

\begin{table*}
\caption{Structure properties and ranking results of the different real networks. Structure properties include network size ($N$), edge number ($E$), degree Heterogeneity ($H=\langle k^2\rangle/\langle k\rangle^2$), degree assortativity ($r$), clustering coefficient ($\langle C \rangle$) and average shortest path length ($\langle d \rangle$).}
\label{tab1}
\begin{center}
\begin{tabular}{p{2.3cm} p{1.4cm} p{1.4cm} p{1.4cm} p{1.4cm} p{1.4cm} p{1.4cm} p{1.4cm} p{1.4cm} p{1.4cm} p{1.4cm} p{1.4cm}}
\hline
\hline
Network &$N$ &$E$ &$H$ &$r$ &$\langle C \rangle$ &$\langle d \rangle$ &$\langle\tau\rangle_{ks}$ &$\langle\tau\rangle_k$ &$\langle\tau\rangle_{MDD}^*$ &$\lambda^*$\\
\hline
Dolphins           &62 &159 &1.327 &-0.044 &0.259  &3.357 &0.563 &0.710 &0.751 &0.62\\
Word               &112 &425 &1.815 &-0.129 &0.173 &2.536 &0.713 &0.803 &0.816 &0.74\\
Jazz               &198 &2742 &1.395 &0.020 &0.618 &2.235 &0.484 &0.526 &0.530 &0.68\\
E. coli            &230 &695 &2.365 &-0.015 &0.224 &3.784 &0.702 &0.683 &0.721 &0.27\\
C. elegans         &297 &2148 &1.801 &-0.163 &0.292 &2.455 &0.614 &0.693 &0.701 &0.66\\
Netsci             &379 &914 &1.663 &-0.082 &0.741  &6.042 &0.453 &0.509 &0.532 &0.59\\
Email              &1133 &5451 &1.942 &0.078 &0.220 &3.606 &0.766 &0.793 &0.809 &0.47\\
TAP                &1373 &6833 &1.644 &0.579 &0.529 &5.224 &0.619 &0.673 &0.688 &0.72\\
Y2H                &1458 &1948 &2.667 &-0.210 &0.071 &6.812 &0.407 &0.428 &0.462 &0.30\\
Power              &4941 &6594 &1.450 &0.004 &0.080 &18.989 &0.348 &0.506 &0.536 &0.55\\
HEP                &5835 &13815 &1.926 &0.185 &0.506 &7.026 &0.535 &0.537 &0.581 &0.38\\
PGP                &10680 &24316 &4.147 &0.238 &0.266 &7.463 &0.457 &0.453 &0.480 &0.24\\
Astro phys         &14845 &119652 &2.820 &0.228 &0.670 &4.847 &0.731 &0.736 &0.753 &0.50\\
Internet           &22963 &48436 &61.978 &-0.198 &0.230 &3.850 &0.554 &0.546 &0.565 &0.15\\
Cond matt          &36458 &171736 &2.960 &0.177 &0.657 &5.476 &0.702 &0.713 &0.743 &0.42\\
\hline
\hline
\end{tabular}
\vspace*{0.0cm}
\end{center}
\end{table*}

\begin{figure}
  \center
  \includegraphics[width=9cm]{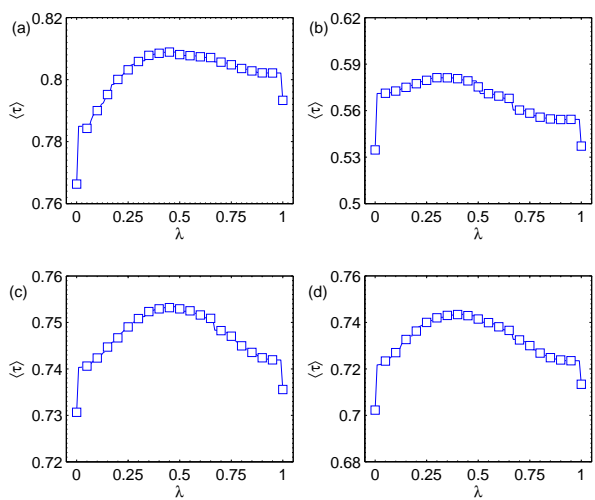}
\caption{(Color online) The value of $\langle\tau\rangle$ of the MDD methods under different parameter $\lambda$. The networks are: (a) email, (b) PGP, (c) astro phys and (d) condensed matter. The results are averaged over $100$ independent realizations. }
\label{fig4}
\end{figure}

\begin{figure}
  \center
  \includegraphics[width=9cm]{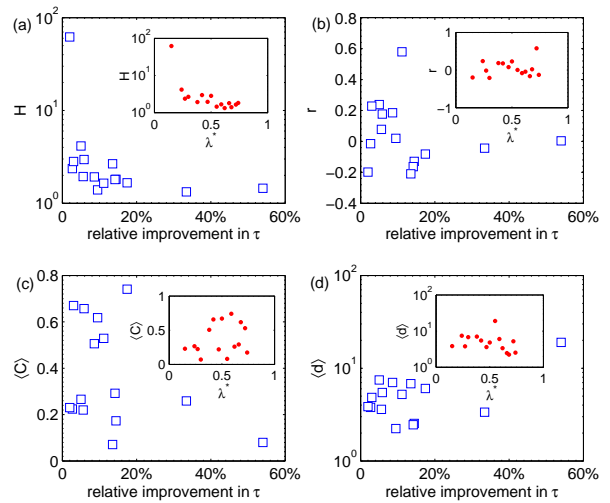}
\caption{(Color online) The relation between the relative improvement in $\tau$ and the network topology parameters including (a) degree heterogeneity, (b) assortativity, (c) cluster coefficient and (d) average shortest path length. The inserts are the relation between the optimal $\lambda^{*}$ and the network topology parameters. Each point in this figure is corresponding to a real network in Table I.}
\label{fig5}
\end{figure}

We further move to investigate how the network structure influences the performance of the MDD method. We first calculate the relative improvement in $\langle\tau\rangle$ as $\frac{\langle\tau\rangle_{MDD}^*-\langle\tau\rangle_{ks}}{\langle\tau\rangle_{ks}}$. The relative improvement can range from $5\%$ to $60\%$ in different networks (the absolute improvement of $\tau$ in some networks can reach as large as $0.188$). Here, we are interested in how the relative improvement get affected by the network topology (mainly including degree heterogeneity, assortativity, average shortest path length and cluster coefficient). The results are shown as scatter plots in Fig. 5 in which each point represents a real network. Obviously, the relative improvement and degree heterogeneity exhibit a negative correlation. This is quite straight forward because the k-shell method is more likely to assign lots of nodes with the same shell when the degree is homogeneous while the MDD method can better distinguish those nodes by considering the exhausted degree. From Fig. 5(b), we can see that the large relative improvement tends to exist in negative assortative region. In the networks with negative assortativity, large degree nodes incline to connect to low degree nodes, in which case some hub connecting many tree-like branches might be formed. Therefore, the MDD method can outperform k-shell method by ranking these hubs higher. As we discussed in the introduction section, k-shell method does not perform well in low clustered networks and it entirely fails in tree networks and BA models. Accordingly, the improvement of MDD method is larger in the networks with low cluster coefficient as shown in Fig. 5(c). Moreover, Fig. 5(d) shows the relation between the relative improvement and the average shortest path length of the networks. The result suggests that MDD method tends to be more effective in networks with large diameter. Since the real networks (especially online social networks) in modern society are extremely large, the MDD will be a suitable method for application.

We recall the results in Fig. 4 in which by increasing (decreasing) a little bit of $\lambda$ when $\lambda=0$ ($\lambda=1$) the $\langle\tau\rangle$ can be significantly improved. This feature suggests that the MDD method is very robust since the MDD can outperform k-shell method and degree method in a large range of $\lambda$. However, we still try to understand better the relation between the optimal $\lambda^{*}$ and the network topology parameter, the results are reported in the insert in Fig. 5. Though the correlation is not clear, we can see some rough trend. When the network are with heterogeneous degree, negative assortativity, low cluster coefficient and large shortest path length, a large $\lambda$ generally performs better. The reason of this trend can be more or less explained based on the above analysis on the relative improvement in $\langle\tau\rangle$.

\section{Conclusion}

The well-known k-shell method is able to identify the most influential spreaders in networks. However, it has two limitations when it is used to rank the nodes spreadability which is defined as the number of final infections resulting from each single given initially-infected node. Specifically, it assigns many nodes with the same rank and it may give a node with strong spreadability with low rank in some cases. We find these limitations are actually due to entirely ignoring the information of exhausted links (i.e., links connecting the remaining nodes to the removed nodes). Accordingly, we propose a Mixed Degree Decomposition (MDD) procedure with a tunable parameter $\lambda$ to rank the spreadability for nodes in networks. By partially considering the exhausted links, we show that the MDD method can significantly improve the ranking accuracy for spreadability and there is an optimal $\lambda$ for each networks. Moreover, the influence of the network structure on the performance of the MDD method is investigated in detail.

Finally, though the MDD method can largely improve the k-shell method in ranking spreaders in complex networks, it is not the optimal way to address this problem. For example, directly considering the number of possible spread pathes and weighting them with some proper damping factor might obtain a more accurate spreadability ranking. However, this method can be with much higher computational complexity than the network decomposition-based methods. Therefore, some more effective and efficient methods are still asked for further investigation.

\section*{acknowledgement}
We would like to thank Yi-Cheng Zhang, Seung-Woo Son, Giulio Cimini and Stanislao Gualdi for helpful suggestions. This work is supported by the Swiss National Science Foundation
(No. 200020-132253).


\begin{thebibliography}{99}
\bibitem{RMP801275} S. N. Dorogovtsev, A. V. Goltsev and J. F. F. Mendes, Rev. Mod. Phys. {\bf 80}, 1275 (2008).
\bibitem{NP3276} V. Colizza, R. Pastor-Satorras and A. Vespignani, Nature Phys. {\bf 3}, 276 (2007).
\bibitem{PRL863200} R. Pastor-Satorras and A. Vespignani, Phys. Rev. Lett. {\bf 86}, 3200 (2001).
\bibitem{PRL93098701} A. E. Motter, Phys. Rev. Lett. {\bf 93}, 098701 (2004).
\bibitem{Nature4641025} S. V. Buldyrev, R. Parshani, G. Paul, H. E. Stanley and S. Havlin, Nature {\bf 464}, 1025 (2010).
\bibitem{EPL8838005} M. Medo, Y.-C. Zhang and T. Zhou, Europhys. Lett. {\bf 88}, 38005 (2009).
\bibitem{PRL89108701} V. M. Eguiluz and K. Klemm, Phys. Rev. Lett. {\bf 89}, 108701 (2002).
\bibitem{PRL90028701} M. Boguna, R. Pastor-Satorras and A. Vespignani, Phys. Rev. Lett. {\bf 90}, 028701 (2003).
\bibitem{PRL97088701} M. A. Serrano and M. Boguna, Phys. Rev. Lett. {\bf 97}, 088701 (2006).
\bibitem{PRL105218701} C. Castellano and R. Pastor-Satorras, Phys. Rev. Lett. {\bf 105}, 218701 (2010).
\bibitem{PRE85015101} M. Schlapfer and L. Buzna, Phys. Rev. E {\bf 85}, 015101(R) (2012).
\bibitem{EPL9518005} A. N. Bishop and I. Shames, Europhys. Lett. {\bf 95}, 18005 (2011).
\bibitem{PRL91247901} R. Cohen, S. Havlin and D. ben-Avraham, Phys. Rev. Lett. {\bf 91}, 247901 (2003).
\bibitem{PRL101058701} Y. Chen, G. Paul, S. Havlin, F. Liljeros and H. E. Stanley, Phys. Rev. Lett. {\bf 101}, 058701 (2008).
\bibitem{NP6888} M. Kitsak, L. K. Gallos, S. Havlin, F. Liljeros, L. Muchnik, H. E. Stanley and H. A. Makse, Nautre Phys. {\bf 6}, 888 (2010).
\bibitem{PhysicaA3911777} D. Chen, L. Lu, M.-S. Shanga, Y.-C. Zhang and T. Zhou, Physica A {\bf 391}, 1777 (2012).
\bibitem{arxiv12020024} R. A. P. da Silva, M. P. Viana and L. da F. Costa, arXiv:1202.0024v1 (2012).
\bibitem{PRE84056105} C. H. Comin and L. da F. Costa, Phys. Rev. E {\bf 84}, 056105 (2011).
\bibitem{PNAS1074491} S. Carmi, S. Havlin, S. Kirkpatrick, Y. Shavitt and E. Shir, Proc. Natl. Acad. Sci. USA {\bf 107},4491 (2010).
\bibitem{PRL103228702} S.-W. Son, B. J. Kim, H. Hong and H. Jeong, Phys. Rev. Lett. {\bf 103}, 228702 (2009).
\bibitem{PRE83045101} A. Zeng, S.-W. Son, C. H. Yeung, Y. Fan and Z. Di, Phys. Rev. E {\bf 83}, 045101(R) (2011).
\bibitem{EPL8748002} A. Zeng, Y. Hu and Z. Di, Europhys. Lett. {\bf 87}, 48002 (2009).
\bibitem{Science286509} A.-L. Barabasi and R. Albert, Science {\bf 286}, 509 (1999).
\bibitem{dolphins} D. Lusseau et al., Behav. Ecol. Sociobiol. {\bf 54}, 396 (2003).
\bibitem{jazz} P. M. Gleiser and L. Danon, Adv. Complex Syst. {\bf 6}, 565 (2003).
\bibitem{netcoauthor_word} M. E. J. Newman, Phys. Rev. E {\bf 74}, 036104 (2006).
\bibitem{email} R. Guimera, L. Danon, A. Diaz-Guilera, F. Giralt and A. Arenas, Phys. Rev. E {\bf 68}, 065103 (2003).
\bibitem{HEP_astrophys_condmatt} M. E. J. Newman, Proc. Natl. Acad. Sci. USA {\bf 98}, 404 (2001).
\bibitem{PGP} M. Boguna, R. Pastor-Satorras, A. Diaz-Guilera and A. Arenas, Phys. Rev. E {\bf 70}, 056122 (2004).
\bibitem{Ecoli} H. Jeong, B. Tombor, R. Albert, Z. N. Oltvai, and A. Barabasi, Nature {\bf 407}, 651 (2000).
\bibitem{Celegans} J. Duch and A. Arenas, Phys. Rev. E {\bf 72}, 027104 (2005).
\bibitem{TAP} A. C. Gavin et al., Nature {\bf 415}, 141 (2002).
\bibitem{Y2H} H. Jeong, S. P. Mason, A. Barabasi, and Z. N. Oltvai, Nature {\bf 411}, 41 (2001).
\bibitem{Power} D. J. Watts and S. H. Strogatz, Nature {\bf 393}, 440 (1998).
\bibitem{Internet} M. E. J. Newman,¡°Networkdata¡±,http://www-personal.umich.edu/¡«mejn/netdata/.

\end{thebibliography}
\end{document}